\documentclass[superscriptaddress,twocolumn,aps,pra,showpacs,
fixfloats]{revtex4}
\usepackage{amsmath}
\usepackage{bm}
\usepackage{dcolumn}
\usepackage{graphicx}

\begin{document}

\title{Confinement and correlation effects in the Xe@C$_{60}$ generalized oscillator strengths}
\author{ M. Ya. \surname{Amusia}}
\affiliation{Racah Institute of Physics, Hebrew University, 91904 Jerusalem, Israel}
\affiliation{A. F. Ioffe Physical-Technical Institute, 194021 St. Petersburg, Russia }
\author{L. V. \surname{Chernysheva}}
\affiliation{A. F. Ioffe Physical-Technical Institute, 194021 St. Petersburg, Russia}
\author{V. K. \surname{Dolmatov}}
\affiliation{Department of Physics and Earth Science, University of North Alabama,
Florence, Alabama 35632, USA}
\date{\today}

\begin{abstract}
The impact of both confinement and electron correlation on
generalized oscillator strengths (GOS's) of endohedral atoms,
$A$@C$_{60}$, is theoretically studied choosing the Xe@C$_{60}$
$4d$, $5s$ and $5p$ fast electron impact ionization as the case
study. Calculations are performed in the
transferred to the atom energy region beyond the $4d$ threshold, 
$\omega =75$--$175$ eV. The calculation methodology combines the plane wave
Born approximation, Hartree-Fock approximation, and random phase
approximation with exchange in the presence of the C$_{60}$
confinement. The confinement is modeled by a spherical
$\delta$-function-like potential as well as by a square well potential to
evaluate the effect of the finite thickness of the C$_{60}$ cage on
the Xe@C$_{60}$ GOS's. Dramatic
distortion of the $4d$, $5p$ and $5s$ GOS's by the confinement is
demonstrated, compared to the free atom. Considerable
contributions of multipolar transitions beyond dipole transitions
in the calculated GOS's is revealed, in some instances. The
vitality of accounting for electron correlation in calculation of
the Xe@C$_{60}$ $5s$ and $5p$ GOS's is shown.
\end{abstract}

\maketitle

\section{Introduction}

Nano-objects $A$@C$_{n}$, consisting of an atom $A$ encapsulated inside the
hollow inner space of a carbon cage C$_{n}$, known as endohedral fullerenes, or
endohedral atoms, or, simply, endohedrals, or confined atoms, have attracted much attention
of investigators. This is because  of their importance to
various basic and applied sciences and technologies. To name a few, one
could emphasize their significance for astrophysics \cite{Astro}, invention of
quantum computers \cite{Harneit}, development of unique superconductors
\cite{SuperC1,SuperC2}, cancer therapy \cite{CancerT}, etc. Understanding of
their quantum structure as well as interaction with various incoming beams
of particles - photons, electrons, ions, etc. - is imperative.
From a theoretical side, the
problem is formidable in complexity due to its multi-faceted
nature. A unique theory that solves this problem once and for all
is yet to be developed. Meanwhile, with the help of simpler,
physically transparent theoretical models, theorists have been
unraveling most unusual aspects of $A$@C$_{n}$ confined atoms,
thereby identifying the most useful experimental studies, which
could be performed. Much of attention has been turned to various
aspects of photo-ionization of
endohedral atoms. The interested reader is referred to review papers
\cite{RPC04,AQC09} as well as some recent papers
\cite{AmC60240'09,Rel&C60'09,AmJETP09,HimadriXe@C60'10,Ludlow2Photo10,BaltOffC60'10}
 and references therein in
addition to other references presented in this paper, for a detailed
introduction into the subject. Many important insights into $A$@C$_{60}$
photoionization have been obtained on the basis of the $\Delta $-potential
\cite{RPC04,AQC09} and $\delta $-potential \cite{Krak98,Balt99} models. In
the $\delta$-potential model the C$_{60}$ cage is assumed to have the zero
thickness and is modeled by a spherical $\delta$-function potential
$V(r)=U_{0}\delta (r-R_{0})$ of an inner radius $R_{0}$ and depth $U_{0}$. In
contrast, the $\Delta $-potential model accounts for the finite thickness
$\Delta $ of the C$_{60}$ cage. It models the cage by a square well potential
of the width $\Delta $. One of spectacular findings, obtained on the basis
of these models, has been the discovery of resonances, termed confinement
resonances (CR's)\cite{Balt99,ConnDolmMans} and correlation confinement
resonances (CCR's) \cite{CCRs}, in the photoionization spectrum of an
endohedral atom. CR's (also referred to as \textit{ordinary} CR's in this
paper) occur in photoionization spectra of endohedral atoms due to
interference of the photoelectron waves emerging directly from the confined
atom $A$, and those scattered off the C$_{60}$ carbon cage. CCR's differ
from these ordinary CR's in that they occur in the spectrum of an
\textit{outer} subshell of the confined atom $A$ due to interference of transitions
from this subshell with ordinary CR's emerging in \textit{inner} shell
transitions, via interchannel coupling \cite{CCRs}. CCR's represent a
novel class of resonances that can exist neither without confinement nor
electron correlation. Both, ordinary CR's and CCR's have attracted much
interest of researchers. In particular, of great importance were
theoretical predictions of a dramatic distortion of the atomic Xe $4d$ giant
resonance by CR's in the Xe@C$_{60}$ $4d$ photoionization made on the
basis of the $\delta$-potential model \cite{AmusiaXe@C60}, $\Delta$-potential model \cite{CCRs},
and time-dependent local density approximation
(TDLDA) \cite{HimadriXe@C60'10} calculations.
 This has stimulated a photoionization experiment that led to a recent experimental
discovery of CR's in the Xe@C$_{60}^{+}$ $4d$ photoionization spectrum \cite{KilcoyneXe@C60}.
The results obtained were in a much better agreement with the $\delta$-potential model
calculated data \cite{AmusiaXe@C60} than with those obtained in the framework of the
$\Delta$-potential model or TDLDA. (Note, beforehand, that for this reason primarily the
$\delta$-potential model is employed in the present paper study).

In contrast to photoionization cross sections of endohedral atoms, little is
known about their generalized oscillator strengths (GOS's). GOS's reflect
the atomic response to fast electron impact ionization.  They are
more complicated and informative parameters than corresponding
photoionization cross sections. This is because, generally, many more multipolar
transitions contribute to electron impact ionization of atoms versus
primarily dipole transition contributions to the photoionization process,
see, e.g., Ref.\ \cite{AmusiaATOM}. The electron spectroscopy of quantum objects
thus serves as another powerful tool for the
study of their structures. However, to date, GOS's of endohedral
 atoms were
investigated only in the theoretical work \cite{BaltGOS09}, and there has been no
associated experimental studies performed. In Ref.\ \cite{BaltGOS09}, the ionization of the
innermost $1s$ subshells of endohedral He@C$_{60}$ and Ne@C$_{60}$ by fast
electrons was chosen as the case study. Both the $\delta$- and
$\Delta$-potential models were employed in the study. Electron correlation was not
accounted for. It was shown that, much as due to photoionization, noticeable ordinary CR's
emerge in GOS's of endohedral atoms as well.

The above mentioned finding on GOS's of endohedral atoms is the only known
to date confinement-related-feature of their electron impact ionization. The knowledge on a possible
significance of electron correlation in GOS's of $A$@C$_{60}$ atoms
is ultimately absent, despite its obvious importance. The present
paper fills in this vacancy in one's knowledge. It advances the initial
understanding of GOS's of such atoms by accounting for a mutual impact of
confinement and correlation on the ionization process. With the general
impetus of the successful experimental study of CR's in the Xe@C$_{60}$ $4d$
photoionization \cite{KilcoyneXe@C60}, we, too, choose to explore the
Xe@C$_{60}$, as the case study. We focus on the $4d$, $5p$ and $5s$ GOS's upon
fast electron impact ionization of the confined Xe assuming that the
transferred to the atom energy exceeds the $4d$ threshold which is
approximately $74$ eV. This is in order to avoid dealing with the C$_{60}$ dynamical
polarization impact on an $A$@C$_{60}$ atom, which is known to be strong at
lower energies \cite{SolovyovC60Screen,AmBalt06,AmusiaJETP}, to simplify
matters. Confinement effects are then accounted for primarily in the framework of the
$\delta $-potential model without regard for said dynamical
polarization of C$_{60}$. The $\Delta $-potential model is employed in some instances as well, to evaluate
the finite-potential-width-impact on the Xe@C$_{60}$ GOS's. In the calculations, the plane wave Born approximation (PWBA) is used
for the fast incoming and scattered electrons. A Hartree-Fock (HF)
approximation in the presence of the C$_{60}$ confinement is employed
relative to the confined atom itself. This completes GOS's calculations in the
zero-order approximation (omitting correlation). Electron correlation is
then accounted for in the framework of the random phase approximation with
exchange (RPAE) \cite{AmusiaATOM,AmusiaPhotoeffect}, as the final step in
the GOS's study.

\section{Theory}

In this section, we provide the outline of the general approach to calculations of GOS's of free ($A$)
and endohedral $A$@C$_{60}$ atoms.

In PWBA, GOS of an $nl$ atomic subshell, $f_{nl}(q,\omega)$, is defined by
[in atomic units (a.u.), but with the energy being measured in Rydbergs (Ry)]
\cite{AmusiaATOM}

\begin{equation}
f_{nl}(q,\omega)=\frac{2(2\lambda +1)N_{nl}\omega }{q^{2}(2l+1)}
\sum_{l^{\prime }\lambda }|Q_{nl,\epsilon ^{\prime }l^{\prime }}^{\lambda
}(q)|^{2}.
\end{equation}
Here, $N_{nl}$ is the number of electrons in the ionizing atomic subshell
$nl $, $q$ is the magnitude of the transferred  linear momentum to the atom upon the collision,
$\omega $ and $\lambda$ are the corresponding transferred energy and orbital momentum, respectively,
$Q_{nl,\epsilon ^{\prime }l^{\prime }}^{\lambda }(q)$ is a reduced matrix
element for the ionization amplitude (in length-form), $\epsilon ^{\prime }$ is the energy of an ejected
electron ($\epsilon^{\prime }=\omega -I_{nl}$, $I_{nl}$ being the $nl$
subshell ionization potential).

In a HF approximation,
\begin{eqnarray}
Q_{nl,\epsilon ^{\prime }l^{\prime }}^{\lambda (\mathrm{HF})}(q) &=&\sqrt{
(2l^{\prime }+1)(2\lambda +1)}\left(
\begin{array}{ccc}
l & l^{\prime } & \lambda \nonumber \\
0 & 0 & 0
\end{array}
\right)  \label{Q} \\
&&\times \int_{0}^{\infty }P_{nl}(r)j_{\lambda }(qr)P_{\epsilon ^{\prime
}l^{\prime }}(r)dr.
\end{eqnarray}
Here, $P_{nl}(r)$ and $P_{\epsilon ^{\prime }l^{\prime }}(r)$ are radial
parts of corresponding HF atomic wave-functions of the initial and final
states of the atom, and $j_{\lambda }(qr)$ is the spherical
Bessel function.

In RPAE, the equation for the GOS reduced matrix element $Q_{nl,\epsilon
^{\prime }l^{\prime }}^{\lambda }(q)$ is more complicated due to the
specific accounting for intershell coupling of the $nl$ $\rightarrow$
$\epsilon l^{\prime }$ transition with electronic transitions from other
subshells of the atom, see Eq.\ ($10.14$) in Ref.\ \cite{AmusiaATOM}:

\begin{eqnarray}
Q_{nl,\epsilon ^{\prime }l^{\prime }}^{\lambda (\mathrm{RPAE})}(q)
&=&Q_{nl,\epsilon ^{\prime }l^{\prime }}^{\lambda (\mathrm{HF})}(q) \nonumber \\
&&+\left( \sum_{
\substack
{
k^{\prime \prime
}l^{\prime \prime \prime }>F, \\
k^{\prime }l^{\prime \prime }\leq F}
}-\sum_{
\substack
{
k^{\prime }l^{\prime \prime
}>F, \\ k^{\prime \prime }l^{\prime \prime \prime }\leq F}
} \right) \nonumber \\
&&\times \frac{U_{nl,\epsilon ^{\prime }l^{\prime };k^{\prime }l^{\prime \prime
},k^{\prime \prime }l^{\prime \prime \prime }}Q_{k^{\prime }l^{\prime \prime
},k^{\prime \prime }l^{\prime \prime \prime }}^{\lambda (\mathrm{RPAE})}(q)}{
\omega -\epsilon _{k^{\prime \prime }l^{\prime \prime \prime }}+\epsilon
_{k^{\prime }l^{\prime \prime }}+i\eta }.
\end{eqnarray}
Here $kl\leq F$ denotes summation over all
occupied atomic states, $kl>F$
marks summation over discrete excited states including integration over continuous spectrum
with the assumption of $\eta \rightarrow +0$,
$\epsilon _{kl}$'s  are the HF energies of corresponding
vacant or occupied atomic states,
$U_{nl,\epsilon ^{\prime }l^{\prime };n^{\prime
}l^{\prime \prime },k^{\prime }l^{\prime \prime \prime }}
= (nl,\epsilon ^{\prime }l^{\prime }|V|n^{\prime
}l^{\prime \prime },k^{\prime }l^{\prime \prime \prime}) -
(nl,\epsilon^{\prime}l^{\prime}|V|k^{\prime}l^{\prime \prime \prime},n^{\prime}l^{\prime \prime})$
is the difference between direct and exchange Coulomb matrix elements of intershell interaction,
respectively. The interested reader is referred to Ref.\ \cite{AmusiaATOM} for more details of
the RPAE methodology.

We now turn to the description of GOS's of $A$@C$_{60}$ endohedral atoms.

Let us first employ the $\delta $-potential model \cite{Krak98,Balt99} to account for the C$_{60}$ confining cage.
This model exploits the following
two key assumptions. First, it is assumed that the size of a confined atom
is much smaller than the C$_{60}$ radius R$_{0}$,
$R_{0}=6.64$ a.u. \cite{R0EA}. This allows one to equal the ground state energies and
electronic wave-functions of the confined atom to those of the
free atom. Second, it is assumed that the thickness $\Delta $ of
the C$_{60}$ cage is much smaller than the wavelength of the
outgoing electron released upon ionization of the confined atom. Hence, the
thickness can be disregarded at all, i.e., $\Delta $ $\rightarrow$ $0$, to a good approximation.
Correspondingly, one can model the C$_{60}$ cage by the $\delta $-function-like potential $U_{\delta}(r)$:
\begin{eqnarray}
U_{\delta}(r)=-U_{\delta}^{(0)}\delta (r-R_{0}).
\label{Udelta}
\end{eqnarray}
Here, $U_{\delta}^{(0)}=0.442$ a.u.\ is the potential depth which was found \cite{Krak98,Balt99} by matching
the calculated electron affinity ($EA$) of C$_{60}$ to the
known one, $EA=2.65$ eV \cite{R0EA}.
The confinement brought impact on the ionization process is then associated
only with modification of the wave-function $P_{\epsilon l^{\prime }}(r)$
of an outgoing electron due to its scattering off the confining cage.
The modification results \cite{Balt99} in $P_{\epsilon l^{\prime }}(r)$ which differs from
that of the free atom $P_{\epsilon l^{\prime }}^{\mathrm{free}}(r)$ only by
a multiplicative factor $D_{l^{\prime }}(k)$ ($k$ being the momentum of an outgoing electron):

\begin{eqnarray}
P_{\epsilon l^{\prime }}(r)=D_{l^{\prime }}(k)P_{\epsilon
l^{\prime }}^{\mathrm{free}}(r),
\label{P}
\end{eqnarray}
where
\begin{eqnarray}
D_{l^{\prime }}(k)=\cos \eta _{kl^{\prime }}\left[ 1-\tan \eta _{kl^{\prime
}}\frac{G_{kl^{\prime }}(R_{0})}{P_{kl^{\prime }}(R_{0})}\right]
\label{Dlk}.
\end{eqnarray}
Here, $G_{kl^{\prime }}$ is the irregular-at-zero solution of the HF equation
for the isolated atom, whereas $\eta _{kl^{\prime }}$ is the additional to the free
atom phase shift due to the $\delta $-potential well:
\begin{eqnarray}
\tan \eta _{kl^{\prime }}(k\text{r})=\frac{P_{kl^{\prime }}^{2}(R_{0})}{
P_{kl^{\prime }}(R_{0})G_{kl^{\prime }}(R_{0})+k/2B}.
\end{eqnarray}

With the help of Eq.\ (\ref{P}), the HF matrix element for the confined atom
GOS amplitude, labeled as $Q_{nl,\epsilon ^{\prime }l^{\prime }}^{\lambda
(\mathrm{\delta HF})}(q)$, differs \cite{BaltGOS09} from that of the free atom, $Q_{nl,l^{\prime }}^{\lambda
(\mathrm{free})}(q)$, Eq.~(\ref{Q}), only by the factor $D_{kl^{\prime }}$:
\begin{eqnarray}
Q_{nl,\epsilon ^{\prime }l^{\prime }}^{\lambda (\delta
\mathrm{HF})}(q)=D_{l^{\prime }}(k)Q_{nl,\epsilon ^{\prime }l^{\prime }}^{\lambda
(\mathrm{free})}(q).
\label{@QHF}
\end{eqnarray}
We will be referring to the described HF approximation for calculating
GOS's of confined atoms as the $\delta$HF approximation; the symbol $\delta
$ emphasizes that the approximation employs the $\delta$-potential concept.

As follows from Eq.~(\ref{Dlk}), the coefficient $D_{l}(k)$ has an
oscillatory character versus $k$ (and, hence, versus the transferred to the
atom energy $\omega $). Therefore, there are resonances - confinement
resonances - emerging in the transition matrix elements for $A$@C$_{60}$
atoms. They translate into resonances either in their photoionization cross
sections or, what is more important to us, generalized oscillator strengths
\cite{BaltGOS09}.

In the framework of the alternative $\Delta$-potential model, the potential $U_{\delta}(r)$, Eq.~(\ref{Udelta}),
is replaced by a short-range square-well potential $U_{\Delta}(r)$ of the width $\Delta$ and depth $U_{\Delta}^{(0)}$:

\begin{eqnarray}
U_{\Delta}(r)=\left\{
\begin{array}{cc}
-U_{\Delta}^{(0)}, & \mbox{at $R_{0}-\frac{1}{2}\Delta \le r \le R_{0}+\frac{1}{2}\Delta$}  \\
0, & \mbox{ otherwise}
\end{array}
\right.
\label{eqDelta}
\end{eqnarray}
Now, excited wavefunctions of the confined atoms $P_{\epsilon l^{\prime }}^{\Delta}(r)$ are not
proportional to wavefunctions $P_{\epsilon l^{\prime }}^{\mathrm{free}}(r)$ of the free atom, Eq.~(\ref{P}).
Instead, the new $P_{\epsilon l^{\prime }}^{\Delta}(r)$ are to be found by a straightforward solution of a ``confined'' HF equation (refereed to as the $\Delta$HF approximation),
i.e., the HF equation which includes the potential
$U_{\Delta}(r)$ in addition to the free atom potential.
 In the present work, we employ the values of $U_{\Delta}^{(0)} = 0.422$ a.u.\ and $\Delta = 1.25$ a.u.\ since they were proven \cite{Dolm11} to result in the best possible
 $\Delta$-model description of experimentally
observed CR's in the Xe@C$_{60}^{+}$ $4d$ photoionization cross section \cite{KilcoyneXe@C60}.
Corresponding HF GOS's amplitudes
 $Q_{nl,\epsilon ^{\prime}l^{\prime }}^{\lambda (\Delta\mathrm{HF})}(q)$ are then calculated with the help of the thus found wavefunctions $P_{\epsilon l^{\prime }}^{\Delta}(r)$.

To account for RPAE electron correlation in GOS's of $A$@C$_{60}$ atoms in either of the discussed models,
the standard free-atom-RPAE-equation is turned into the ``confined-atom-RPAE-equation'' by a straightforward replacement of all
free atomic excited state wavefunctions $P_{\epsilon l'}^{\rm free}(r)$ by the above discussed wavefunctions $P_{\epsilon l'}^{\delta}(r)$ or $P_{\epsilon l'}^{\Delta}(r)$, respectively.
Similar to the used $\delta$HF and $\Delta$HF abbreviations, we refer to such confined-atom-RPAE methodology as $\delta$RPAE and $\Delta$RPAE, respectively.

In particular, the $\delta$RPAE equation transforms into
\begin{eqnarray}
Q_{nl,\epsilon ^{\prime }l^{\prime }}^{\lambda (\delta\mathrm{RPAE})}(q)
&=&Q_{nl,\epsilon ^{\prime }l^{\prime }}^{\lambda (\delta\mathrm{HF})}(q) \nonumber \\
&&+\left( \sum_
{
\substack
{
k^{\prime }l^{\prime \prime }\leq F,  \\ k^{\prime
\prime }l^{\prime \prime \prime }>F
}
}
D_{l^{\prime \prime \prime
}}^{2}(k^{\prime \prime })-\sum_{\substack{ k^{\prime }l^{\prime \prime }>F,
\\ k^{\prime \prime }l^{\prime \prime \prime }\leq F}}D_{l^{^{\prime \prime
}}}^{2}(k^{\prime })\right) \nonumber \\
&&\times\frac{U_{nl,\epsilon ^{\prime }l^{\prime };k^{\prime }l^{\prime \prime
},k^{\prime \prime }l^{\prime \prime \prime }}Q_{k^{\prime }l^{\prime \prime
},k^{\prime \prime }l^{\prime \prime \prime }}^{\lambda (\delta \mathrm{RPAE})
}(q)}{\omega -\epsilon _{k^{\prime \prime }l^{\prime \prime \prime
}}+\epsilon _{k^{\prime }l^{\prime \prime }}+i\eta }.
\label{@RPAE}
\end{eqnarray}
The intershell interaction term in the $\delta$RPAE equation (the second term on the right-hand-side of the equation)
explicitly depends on the oscillatory parameter $D_{l}(k)$. Hence,
the $\delta$RPAE approximation is capable of accounting for CCR's in the GOS's spectra. It accounts for ordinary CR's as well,
owing to the term $Q_{nl,\epsilon ^{\prime }l^{\prime }}^{\lambda (\delta\mathrm{HF})}(q)$ (the first term on the right-hand-side of
the $\delta$RPAE equation) which itself is determined by Eq.\ (\ref{@QHF}).

As for the alternative $\Delta$RPAE
equation, the latter accounts for CR's and CCR's implicitly, via final-state and intermediate-state functions $P_{\epsilon l}^{\Delta}(r)$.

\section{Results and discussion}

\begin{figure}[tbp]
\includegraphics[width=7cm]{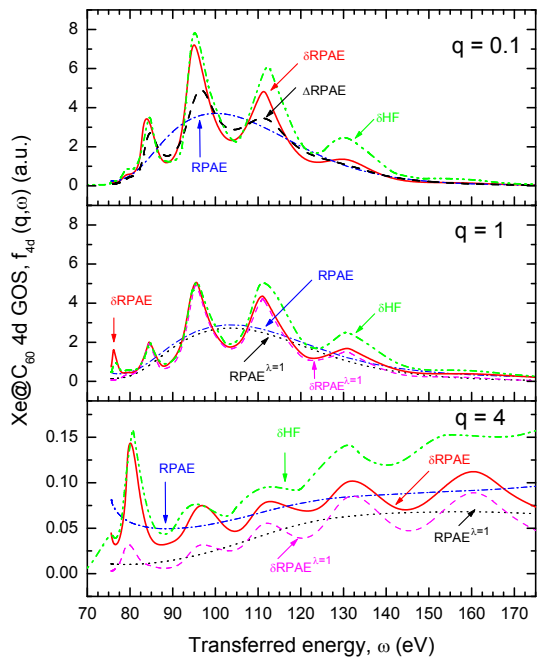}
\caption{(Color online) The $4d$ generalized oscillator strengths $f_{4d}(q,
\protect\omega)=\sum_{\protect\lambda}f_{4d}^{\protect\lambda}(q,\protect
\omega)$ upon fast electron impact ionization of Xe@C$_{60}$ ($\protect
\delta $HF and $\protect\delta$RPAE) and free Xe (HF and RPAE). Calculated results marked as RPAE$^{\protect\lambda=1}$ and
$\protect\delta$RPAE$^{\protect\lambda=1}$ relate to $f_{4d}^
{\protect\lambda=1}(q,\protect\omega)$ due to only dipole transition contributions.}
\label{e-Xe4d}
\end{figure}

Our $\delta $HF, $\Delta $HF, $\delta $RPAE and $\Delta $RPAE
calculated results for the $4d$ GOS's $f_{4d}(q,\omega )$ of
Xe@C$_{60}$, upon fast electron impact ionization, are presented
in Fig.~\ref{e-Xe4d} along with corresponding HF and RPAE
calculated data for free Xe. Calculations were
performed for transferred momenta $q=0.1$, $1$, and $4$
and accounted for major monopole ($\lambda
=0$), dipole ($\lambda =1$), quadrupole ($\lambda =3$), and octupole
($\lambda =3$) multipolar contributions to $f_{4d}(q,\omega )$.
RPAE and $\delta $RPAE calculations included dominant intershell interaction
between transitions from $4d$, $5s$, and $5p$ subshells.
 A number of spectacular trends in $f_{4d}(q,\omega )$ versus $\omega $, $q$,
and $\lambda $ is seen.

One of the trends is the presence of strong oscillations  in the $f_{4d}(q,\omega )$ of Xe@C$_{60}$.
They are absent in the $4d$ GOS's of free atom. Thus, the oscillations are due to the C$_{60}$ confinement,
thereby featuring the emergence of confinement resonances in the $4d$ GOS's of Xe@C$_{60}$.

Furthermore,
interesting, relative intensities and positions of these CR's appear to be noticeably changing
 with increasing value of $q$ in the whole energy region, including near threshold. As a results, the calculated
$f_{4d}(q=0.1,\omega )$ and $f_{4d}(q=1,\omega )$ have little to do with
$f_{4d}(q=4,\omega )$.
Moreover, this impact also results in that the GOS's of free Xe and, on the average, of
Xe@C$_{60}$ are found to exhibit a strong, broad resonance for  $q=0.1$ and $q=1$ in
contrast to $q=4$, in the energy region under discussion. To understand this,
we also plotted in Fig.~\ref{e-Xe4d} calculated data of a trial $\delta$RPAE calculation [labeled as  $f_{4d}^{\lambda =1}(q,\omega )$]
accounting only for dipole contributions to the $4d$ GOS's.
The comparison of the total GOS $f_{4d}(q,\omega )$ with $f_{4d}^{\lambda =1}(q,\omega )$ shows
the dominant role of dipole transitions in the ionization process for smaller $q$'s, $q=0.1$
and $1$, both for free Xe and Xe@C$_{60}$ [note, for
$q=0.1$ dipole channels exceed other channels by several orders of
magnitude, for which reason $f_{4d}(q,\omega )$ is undistinguished
from $f_{4d}^{\lambda =1}(q,\omega )$].
 The Xe dipole $4d$
$\rightarrow $ $\epsilon f$ transition is known to exhibit a strong resonance
versus $\omega $, known as the $4d$ giant resonance
\cite{AmusiaPhotoeffect}. Consequently, the nature of the strong, broad resonance seen in the $4d$
GOS's at $q=0.1$ and $1$ is the same as in the case of the Xe $4d$
photoionization, i.e., it is the familiar $4d$ dipole giant resonance. For a larger
value of $q$, $q=4$, other channels beyond the dipole channel acquire
considerable strengths compared to the dipole channel. This can be judged by
comparing total $f_{4d}(q,\omega )$ with partial $f_{4d}^{\lambda =1}(q,\omega )$ for
$q=4$ displayed in Fig.~\ref{e-Xe4d}. This explains why $f_{4d}(q=4,\omega )$ has, in general,
little in common with the $4d$ GOS's for smaller $q$'s, as well as why $f_{4d}(q=4,\omega)$
does not exhibit the $4d$ giant resonance - dipole transitions matter little.
Thus, the transferred momentum as well as multipolar impacts on the $4d$ GOS's of both free Xe and Xe@C$_{60}$
are found to be considerable.

Moreover, Fig.~\ref{e-Xe4d} additionally features a varying role of multipolar contributions to GOS's near threshold. A trial
calculation showed that the first resonance maximum in $4d$ GOS's near threshold
is primarily due to monopole channels for $q=1$, whereas it is mainly due to
octupole channels for a larger $q=4$.

Finally, it also obvious from Fig.~\ref{e-Xe4d} that the alternative $\Delta$RPAE calculation
of the Xe@C$_{60}$ $4d$ GOS results in lower and yet clearly prominent intensities of emerged CR's
compared to the $\delta$RPAE data; this is in line with results of the previous theoretical study \cite{BaltGOS09}.
Note, since GOS's of endohedral atoms have been experimentally unexplored, the question of which of the
used models is most appropriate remains open.

\begin{figure}[tbp]
\includegraphics[width=7cm]{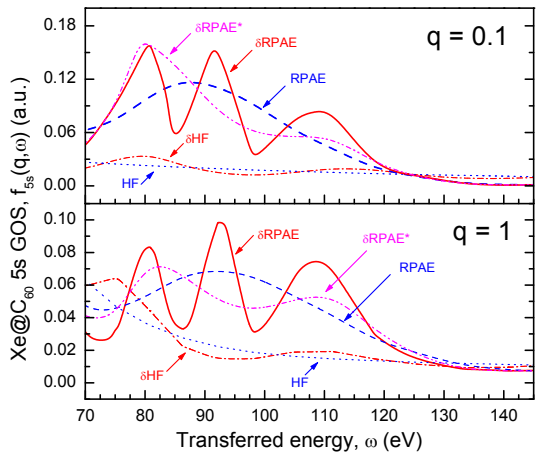}
\caption{(Color online) Calculated data for generalized oscillator strengths
$f_{5s}(q,\protect\omega )$ of free Xe (HF and RPAE) and Xe@C$_{60}$
($\protect\delta$HF and $\protect\delta$RPAE).
${\protect\delta}$RPAE* labels the fictitious calculated data for the Xe@C$_{60}$ $5s$
GOS's (see text).}
\label{e-Xe5s}
\end{figure}

We now proceed to the discussion of the Xe and Xe@C$_{60}$ $5s$ GOS's. Corresponding calculated
HF and RPAE (for free Xe) as well as $\delta$HF and $\delta$RPAE
data for $f_{5s}(q,\omega )=\sum_{\lambda }f_{4d}^{\lambda }(q,\omega )$ are
depicted in Fig.~\ref{e-Xe5s} for $q=0.1$ and $1$ (the $5s$ GOS's
for $q=4$ appear to be negligible compared to those for $q=0.1$ or
$1$, thus presenting little interest for discussion). As in the
above study, calculations accounted for contributions of major
monopole, dipole, quadrupole, and octupole ionization channels to
$f_{5s}(q,\omega )$ as well as intershell coupling between
transitions from the Xe $4d$, $5s$, and $5p$ subshells both in RPAE and
$\delta$RPAE equations.
The calculated data show that, as in the known case of the Xe $5s$ photoionization
\cite{AmusiaPhotoeffect}, the $5s$ GOS's of free Xe are ultimately affected by electron correlation, both for smaller and bigger values of $q$.
This clearly follows from the comparison of
corresponding HF and RPAE calculated data. The same tendency is found to preserve
in endohedral Xe@C$_{60}$ as well, cf.\ $\delta$HF and $\delta$RPAE calculated data.
Furthermore, similar to $4d$ GOS's, the Xe@C$_{60}$ $5s$ GOS's are found to be dramatically distorted
by confinement. Indeed, the presence of the C$_{60}$ confinement results in the emergence of three strong oscillations (resonances)
in $f_{5s}(q,\omega )$ at given $\omega$'s, cf.\ RPAE and $\delta$RPAE calculated data.
 Noting that
the resonance positions are about the same as the position of ordinary CR's in the $4d$ GOS's at
approximately $83$, $95$, and $102$ eV, one might be tempted to interpret the resonances in $5s$ GOS's
as CCR's, i.e., being induced in $f_{5s}(q,\omega )$ by the three ordinary CR's in $4d$ ionization
channels, via intershell interaction, as in the case of the Xe@C$_{60}$ 5s photoionization \cite{CCRs}. This, however,
would not be entirely correct. Indeed, e.g., such interpretation would fail to explain why a weak CR in
$f_{4d}(q,\omega)$ at about $83$ eV induces as strong CCR in $f_{5s}(q,\omega)$ as the two other
stronger neighboring resonances. The actual origin of the three emerged resonances in $5s$ GOS's is thus intriguing.

 To unravel the nature of the three resonances in question in $f_{5s}(q,\omega)$ of Xe@C$_{60}$, we
 performed a fictitious trial
$\delta$RPAE calculation of the $5s$ GOS's. There, we artificially eliminated ordinary CR's
from corresponding coupling $4d$ ionization channels in $\delta $RPAE, Eq.\ (\ref{@RPAE}).
 This was achieved by substituting the
excited wavefunctions of the $4d$ electrons of \textit{free} Xe instead of those of
Xe@C$_{60}$ into Eq.\ (\ref{@RPAE}).
This fictitious methodology will be referred to/labeled as $\delta$RPAE*. Corresponding
$\delta$RPAE* calculated data are depicted in Fig.~\ref{e-Xe5s} as well. One
interesting important observation is that $\delta$RPAE* intershell interaction noticeably increases CR's presented
in $\delta$HF calculated  $f_{5s}(q,\omega)$ for both values of $q$ and also shifts a lower energy CR in
$f_{5s}(q=1,\omega)$ from about $75$ to about $83$ eV. Such $\delta$RPAE* calculation, however, does not bring a third (middle) resonance
in $f_{5s}(q,\omega)$. This resonance emerges only in a true  $\delta$RPAE calculation (which accounts for CR's in $4d$ ionization channels) of
$f_{5s}(q,\omega)$, and its position coincides with that of a middle resonance in $4d$ GOS's. Thus, our first conclusion in unraveling
the nature of the three resonances in $f_{5s}(q,\omega)$ is that the middle resonance is undoubtedly correlation confinement resonance, CCR,
by nature [it does not exist without simultaneous impact of intershell interaction and confinement on $f_{5s}(q,\omega)$]. What about the left and right
resonances in $f_{5s}(q,\omega)$? Their nature is more complicated. Qualitatively, they exist in $f_{5s}(q,\omega)$ even without coupling with CR's in
$4d$ ionization channels (see $\delta$RPAE* calculated data), i.e., the resonances uncovered by the performed $\delta$RPAE* calculation are ordinary CR's.
 In a rare, unique occasion,  these CR's (see $\delta$RPAE* calculated data) peak at about the same
energies as the left and right \textit{ordinary} CR's in $4d$ GOS's, Fig.~\ref{e-Xe4d}. Therefore, when the $4d$ CR's are accounted
for in a true $\delta$RPAE calculation
they strongly affect the two originally existing $\delta$RPAE* ordinary CR's in $f_{5s}(q,\omega)$, thereby enhancing them, via intershell interaction.
Thus, the left and right resonances in true  $\delta$RPAE calculated
data for $f_{5s}(q,\omega)$ are the result of intershell coupling of the ordinary CR's in $4d$ channels with \textit{existing} ordinary
 CR's in $5s$ ionization channels.  Therefore,
the left and right resonances in $f_{5s}(q,\omega)$ are neither purely ordinary CR's nor purely CCR's. Rather, they may be termed
as \textit{CR-CR-correlation-interference-resonances} - a new type of resonances which have not been met earlier, to the best of our knowledge.
This interpretation, in particular, explains why the lower left resonance in $f_{5s}(q,\omega)$ is as strong as the middle CCR.
To start with, it was relatively strong from the very beginning
(see $\delta$RPAE* calculated data). Next, the interaction with a weaker left $4d$ CR makes this $5s$ resonance somewhat stronger, so that
it now matches the middle CCR
which is brought about by the strongest middle CR in the $4d$ channel.
To summarize, the discussed  $5s$ GOS resonance spectrum of Xe@C$_{60}$ has neither a purely CR nor CCR nature. Rather, it consists of
one CCR and two CR-CR-correlation-interference-resonances. Note, this makes the $5s$ GOS spectrum of Xe@C$_{60}$ be unique and different
in its origin from corresponding $5s$ Xe@C$_{60}$ photoionization spectrum \cite{CCRs} which consists of purely CCR's.

\begin{figure}[tbp]
\includegraphics[width=7cm]{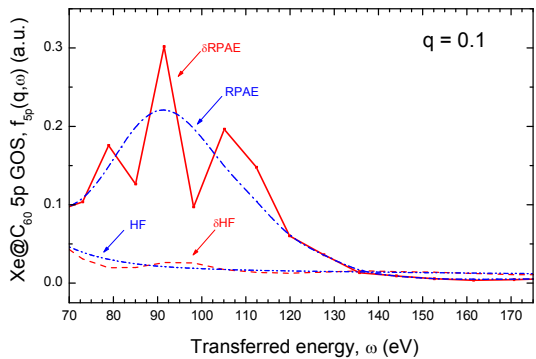}
\caption{(Color online) Calculated data for generalized oscillator strengths
$f_{5p}(q,\protect\omega )$ of free Xe (HF and RPAE) and Xe@C$_{60}$
($\protect\delta$HF and $\protect\delta$RPAE), as marked.}
\label{e-Xe5p}
\end{figure}

Of specific interest are the GOS's of $5p$ electrons in the considered
 $\omega $ region. Since it is well above the $5p$ ionization threshold,
 $f_{5p}(q,\omega )$ for Xe@C$_{60}$ could have been expected to have negligible or
 very weak, at best, CR's.  This seems to be in
line with a theory of scattering of particles off a potential
well or barrier. Indeed, at a sufficiently high energy
of the outgoing electron, the corresponding coefficient
of reflection off a finite potential well or barrier is small. As a result, the interference
effect between the outgoing and scattered electron
waves becomes weak, and so are the associated CR's. This, however, is true only in terms of an
independent particle approximation. As was shown in Ref.~\cite{ResurrectedCRs}, where photoionization
of endofullerenes was chosen as a case study,
CR's can reappear - resurrect - and be strong at the transferred to the atom energy far exceeding (by thousands of eV)
the $nl$ ionization threshold, as a general phenomenon.
This will happen at transferred energies, which correspond to opening of inner-shell
photoionization channels, whose intensities exceed by far the
intensity of transitions from the outer subshell of the confined
atom and which are strongly coupled with the innershell
transitions. This is just the case with the Xe@C$_{60}$ $5p$ ionization above the $4d$ threshold.
Indeed, as known from photoionization studies \cite{AmusiaPhotoeffect},  the Xe $5p$ ionization
is affected strongly by intershell interaction with $4d$ transitions. As was found above,
the Xe@C$_{60}$  $4d$ GOS's are (a) strong and (b) have pronounced CR's, see Fig.~\ref{e-Xe4d}.
Therefore, one can predict the emergence of strong CCR's in the Xe $5p$ GOS when intershell interaction between the
$5p$ and $4d$ ionization amplitudes is accounted for in $\delta$RPAE calculation, as in the above detailed case of $5s$ GOS's.
Corresponding HF, RPAE, $\delta$HF and $\delta$RPAE calculated data for $f_{5p}(q,\omega )$ for $q=0.1$ are depicted
in Fig.~\ref{e-Xe5p}.
The presented results are self-explanatory. In brief, there is the only weak CR related oscillation in the $\delta$HF calculated $5p$ GOS
for $q=0.1$, in contrast to three strong resonance features present in the $\delta$RPAE calculated data. The latter are
found to be induced in the $5p$ GOS by CR's in the $4d$ GOS amplitudes, i.e., they are CCR's. They are strong, in a full accordance with the above made prediction.
In addition, calculated data show that, as in the known case of the free Xe $5p$ photoionization \cite{AmusiaPhotoeffect},
 the $5p$ GOS's of both free Xe and Xe@C$_{60}$ are completely determined by intershell correlation with the $4d$ subshell. This clearly follows
from the comparison between HF and RPAE calculated data on the one hand, and  $\delta$HF and $\delta$RPAE calculated data on the other hand.

\section{Conclusion}

In the present work we focused on the study of the impact of the C$_{60}$
confinement on the $4d$, $5s$ and $5p$ generalized oscillator strengths of
Xe@C$_{60}$, in the energy region above the $4d$ threshold, where, in our
opinion, the most interesting effects occur. We hope that the discovered
impact of the transferred momentum $q$, electron correlation, and
confinement on generalized oscillators strengths of Xe@C$_{60}$ will
challenge experimentalists to verify our predictions. Theorists, we hope,
will be driven by the desire to improve the made predictions with the help
of more sophisticated theories. All this would indisputably result in
uncovering of a richer variety of possible effects outside of the made
predictions, thereby advancing this field of endeavor.

\section{Acknowledgements}
M.Ya.A.\ and L.V.C.\ acknowledge the support received from the Israeli-Russian Grant RFBR-MSTI no.\ 11-02-92484.
V.K.D., acknowledges the NSF support, Grant no.\ PHY-0969386.

\end{document}